\documentclass[10pt,conference]{IEEEtran}

\IEEEoverridecommandlockouts
\usepackage{amsmath,amssymb,amsfonts}
\usepackage[backend=biber,style=ieee,natbib=true]{biblatex} 
\usepackage{algorithmic}
\usepackage{graphicx}
\usepackage{textcomp}
\usepackage{enumitem}
\usepackage{paralist}
\usepackage{booktabs}
\usepackage{mathtools}
\usepackage{multirow}
\usepackage{xcolor}
\usepackage{makecell}
\usepackage{csquotes}
\usepackage{paralist}
\usepackage{graphicx}
\usepackage{amsmath}
\usepackage{mathtools}

\DeclarePairedDelimiter\floor{\lfloor}{\rfloor}

\def\BibTeX{{\rm B\kern-.05em{\sc i\kern-.025em b}\kern-.08em
    T\kern-.1667em\lower.7ex\hbox{E}\kern-.125emX}}

\renewcommand{\bibfont}{\footnotesize} %
\ifCLASSOPTIONcompsoc
    \usepackage[caption=false, font=normalsize, labelfont=sf, textfont=sf]{subfig}
\else
\usepackage[caption=false, font=footnotesize]{subfig}
\fi    
\usepackage{capt-of}

\begin{document}
\newcommand{\myparagraph}[1]{\noindent \textit{#1.}}

\def\numtraces{24,000}

\title{Simple Fault Localization using Execution Traces}

\author{\IEEEauthorblockN{Julian Aron Prenner}
\IEEEauthorblockA{\textit{Free University of Bozen-Bolzano}\\
Bozen-Bolzano, Italy \\
prenner@inf.unibz.it}
\and
\IEEEauthorblockN{Romain Robbes}
\IEEEauthorblockA{\textit{Univ. Bordeaux, CNRS, Bordeaux INP},\\ \textit{LaBRI, UMR 5800} \\
Bordeaux, France \\
romain.robbes@u-bordeaux.fr}
}

\maketitle

\begin{abstract}
Traditional spectrum-based fault localization (SBFL) exploits differences in a program's coverage spectrum when run on passing and failing test cases. However, such runs can provide a wealth of additional information beyond mere coverage. Working with thousands of execution traces of short programs submitted to competitive programming contests and leveraging machine learning and additional runtime, control-flow and lexical features, we present simple ways to improve SBFL. We also propose a simple trick to integrate context information. Our approach outperforms SBFL formulae such as Ochiai on our evaluation set as well as QuixBugs and requires neither a GPU nor any form of advanced program analysis. Existing SBFL solutions could possibly be improved with reasonable effort by adopting some of the proposed ideas. 
\end{abstract}


\section{Introduction}
The goal of fault localization (FL) is to spot bugs in software. This can save development cost and time as well as ease software maintenance. Fault localization is also an important step in automated program repair where repair effort is concentrated on code locations found suspicious (i.e., possibly buggy) by a fault localization system.

A commonly employed fault localization method is spectrum-based fault localization (SBFL). While many different kinds of program spectra exist~\citep{harroldEmpiricalInvestigationProgram1998a}, most of the time, SBFL relies on the coverage or hit spectrum of the target program. This spectrum indicates which program elements are covered, that is executed, by passing or failing test cases. Program elements can comprise statements, lines, blocks or even methods. However, in this work we focus on the line level. From the hit spectrum, four central features are derived \emph{for each line} $l$ (or other program element) of a program: 
\begin{description}
\item[$e_p$,] the number of passing test cases covering line $l$,
\item[$e_f$,] the number of failing test cases covering line $l$,
\item[$n_{p}$,] the number of passing test cases \emph{not} covering line $l$,
\item[$n_{f}$,] the number of failing test cases \emph{not} covering line $l$.
\end{description}

By weighing these four features by means of a risk evaluation formula~\citep{xieTheoreticalAnalysisRisk2013} (also referred to as ranking metric) such as Ochiai~\citep{ochiaiZoogeographicalStudiesSoleoid1957, jonesVisualizationTestInformation2002} each line is assigned a score. Lines can then be ranked by this score (usually high to low) to find the most likely fault locations.

\paragraph*{This work} In this work we combine three simple ideas to improve spectrum-based localization performance. First, we include features from \emph{additional program spectra} (e.g., the count spectrum) that can easily be obtained from execution traces and that do not require any form of advanced static or runtime analysis. Second, we take \emph{context} into account. That is, the calculation of the score for a line $l$ may consider features belonging to surrounding lines (within a sliding window). Finally, instead of simple formulae we follow a \emph{data-driven} approach and use extracted features to fit a gradient boosted machine (GBM). Section~\ref{sec:related-work} contrasts these points with previous work. 

We rely on simple Python programs from the RunBugRun~\citep{prennerRunBugRunExecutableDataset2023} dataset (see Section~\ref{sec:methodology} for details). We also evaluate our model on the Python version of the QuixBugs dataset~\citep{linQuixBugsMultilingualProgram2017} and provide a simple ablation study~(Section \ref{sec:ablation}).

\paragraph*{Results} On both, our evaluation set of RunBugRun bugs as well as on QuixBugs~\citep{linQuixBugsMultilingualProgram2017}, our GBM model outperforms commonly used formulae taken from well-known SBFL systems such as Ochiai~\citep{jonesVisualizationTestInformation2002}, DStar~\citep{wongDStarMethodEffective2014} and Tarantula~\citep{jonesEmpiricalEvaluationTarantula2005}.
Results are detailed in Section~\ref{sec:results}.

\paragraph*{Conclusions}
Our work suggests that SBFL performance can be improved by relatively simple means; in particular, by using a count spectrum instead of a hit spectrum and a rolling window over multiple lines to capture some form of context. These ideas are relatively simple to implement and could be integrated into existing SBFL systems and frameworks. In contrast to other recent deep learning-based approaches, the used GBM models are quick to train, have modest hardware requirements and do not require a GPU. 

\section{Related Work}\label{sec:related-work}

We focus on previous work related to the most important aspects of our approach, that is
\begin{inparaenum}[i)]
\item previous learning-based FL and in particular SBFL methods,
\item previous methods exploiting alternative or multiple spectra,
\item approaches that address the problem of missing context in SBFL and finally
\item methods that rely on traces.
\end{inparaenum}

\paragraph*{Learning-based}
Certainly, the idea of fitting a SBFL score function is not new. \citet{xuanLearningCombineMultiple2014} propose MULTRIC, that combines multiple ranking formulae using learned weights.
\citet{yooEvolvingHumanCompetitive2012} uses genetic programming to evolve an optimal risk evaluation formula. Early experiments with neural networks were done by \citet{wongBpNeuralNetworkbased2009}.
In recent years, deep learning has been successfully applied to the problem of fault localization in various different forms~\citep{liDeepFLIntegratingMultiple2019, liFaultLocalizationCode2021}, including large code language models~\citep{yangLargeLanguageModels2024, jiImpactLargeLanguage2024} and graph neural networks (GNNs)~\citep{rafiBetterGraphNeural2024, louBoostingCoveragebasedFault2021, zhangContextAwareNeuralFault2023, qianAGFLGraphConvolutional2021, qianGNet4FLEffectiveFault2023}.
Our work employs GBMs to predict fault scores on a per-line level.

\paragraph*{Additional spectra \& features}
\citet{yilmazTimeWillTell2008} use the time spectrum (i.e., execution time features) for localization. This work also employs some form of learning, as it uses Gaussian Mixture Models (GMMs) to model time.
\citet{zhangCapturingPropagationInfected2009} rely on a control flow spectrum (edge spectrum) to localize faults. They focus on cases where the effect of a fault may show at a different location than its true cause. \citet{santelicesLightweightFaultlocalizationUsing2009} devise a localization approach that uses def-use pairs and a branch coverage spectrum. 
DEPUTO~\citep{abreuRefiningSpectrumbasedFault2009} improves spectrum-based fault localization through the use of abstract interpretation. FLUCCS~\citep{sohnFLUCCSUsingCode2017} uses code change metrics such as age, churn or complexity to aid localization. 
DeepFL~\citep{liDeepFLIntegratingMultiple2019} combines features from various different localization methods, including SBFL, mutation-based fault localization (MBFL) as well as code complexity and text features.
In this work, in addition to coverage and count spectrum features, we also use very simple control flow features. We also experiment with lexical features.

\paragraph*{Using context}
Traditional SBFL methods determine suspiciousness separately for each program element without taking into account possible dependencies between program elements or other contextual information.
This problem has been identified in previous work~\citep{sarhanSurveyChallengesSpectrumBased2022} and several attempts have been made to tackle it. We should note that context is taken here in a very broad sense and includes several different ways to factor dependencies between program elements or simply multiple program elements into the suspiciousness score calculuation.
\citet{zhaoContextAwareFaultLocalization2011} propose calculating suspiciousness scores for control flow edges. We also experiment with simple control flow features, however, without carrying out any advanced control flow analysis. 
Like our work, \citet{laghariUseSequenceMining2018} use a sliding a window, albeit to mine call sequences from call traces. We use sliding windows and simple control flow features to \enquote{contextualize} per-line features.

While DeepFL~\citep{liDeepFLIntegratingMultiple2019} employs LSTMs, the spacial dimension of the LSTM (often referred to as time dimension) seems not to be used to capture context over multiple lines or statements but to combine features of different types belonging to a single program element.
In contrast, \citet{liFaultLocalizationCode2021} convert the spectrum information into images (or rather a feature map) and use them to train convolutional neural networks (CNNs); here one spatial dimension seems to span multiple program elements. Similarly, \citet{zhangImprovingDeeplearningbasedFault2021} train a series of different deep learning architectures, including CNNs, LSTMs and MLPs, on spectral data where one of the input dimensions can span multiple program statements or lines, thus allowing the network to model some form of basic context. 
CAN~\citet{zhangContextAwareNeuralFault2023} aim to contextualize FL by modeling dependencies between program elements using graphs and GNNs.
Finally, some of the previously mentioned work, for instance, the use of edge spectra~\citep{zhangCapturingPropagationInfected2009} could also be said to capture some form of context.
While we are not using neural networks, our sliding window technique is somewhat reminiscent of convolution, used in CNNs.

\paragraph*{Using traces}
Not much work explicitly mentions the use of execution traces. However, similar types of information may be used implicitly, e.g., for control flow analysis or even to determine coverage. AMPLE~\citep{dallmeierLightweightBugLocalization2005} and SPEQTRA~\citep{laghariLocalisingFaultsTest2015} both use method call traces to locate faults at the class level. Similarly, \citet{heEnhancingSpectrumBasedFault2020} employ tracing to collect not only coverage but also call relation information. SmartFL~\citep{zengFaultLocalizationEfficient2022} uses probabilistic modeling on information which is, in addition to static analysis, also obtained from execution traces.

\section{Methodology}\label{sec:methodology}

Our approach involves five main steps:
\begin{inparaenum}[1)]
\item obtaining execution traces,
\item calculating line-level features,
\item calculating feature windows,
\item training and finally
\item evaluation.
\end{inparaenum}
We detail each step below.

\subsubsection{Obtaining traces}
In our work we target buggy Python programs from the RunBugRun~\citep{prennerRunBugRunExecutableDataset2023} dataset which itself is based on the CodeNet dataset~\citep{puriCodeNetLargeScaleAI2021} and contains programs originally stemming from programming competitions such as AtCoder. We select from RunBugRun a subset of Python bugs with at least one failing and one passing test case. Each bug of said subset is instrumented and executed on all its test cases (the test cases are also part of RunBugRun).
For instrumentation, we rely on Python's \texttt{sys.settrace}\footnote{\url{https://docs.python.org/3/library/sys.html#sys.settrace}} system.
This allows us to install a trace callback that is called whenever the Python interpreter is about to execute a line of program code. We use this callback to collect the following information (for each program step):
\begin{inparaenum}[1)]
\item the number of the line being currently executed,
\item the number of the previously executed line,
\item the current value of \enquote{primitive} local variables (\texttt{int}, \texttt{bool}, \texttt{float}),
\item the length of strings, lists, sets and numpy arrays
\item a high resolution timestamp,
\item a step counter (i.e., number of total executed steps or lines so far).
\end{inparaenum}

In this way, we obtain traces for over \numtraces{} bugs. These traces record up to millions of execution steps and their \emph{compressed} file sizes range from a few kilobytes to several hundreds of megabytes in extreme cases. In order to be suitable as input for a machine learning algorithm this data must first be condensed as described below. Importantly, in this work we only use line number information, that is items 1) and 2); all of our features can be derived from these two data points.

\subsubsection{Calculating line-level features}
To condense traces we aggregate trace records by bug ID and test outcome (pass or fail) and calculate the following \emph{per-line} features. 

\begin{description}
\item[Execution counts and coverage (spectrum)]
We calculate $e_p$, $e_f$ as described earlier, as well as $N_f$ and $N_p$, the total number of passed and failed tests, respectively. Moreover, we also calculate the corresponding count spectrum, that is, not only whether a certain
line $l$ was executed but also how often. We also include normalized versions of these features, i.e., the count spectrum divided by the total number of executions.
Similarly, $p_p$ and $p_f$, pass and fail rate, are the normalized versions of $e_p$ and $e_f$.
Note that all of the above features can be derived from the line number information. For instance, if a specific line $l$ appears in the trace it was hit and is thus covered. Similarly, by counting occurrences of $l$ we can obtain a count spectrum.
\item[SBFL formulae (spectrum)] 
We calculate ranking scores using Ochiai~\citep{jonesVisualizationTestInformation2002}, DStar\textsuperscript{2}~\citep{wongDStarMethodEffective2014} and Tarantula~\citep{jonesEmpiricalEvaluationTarantula2005} as follows and use each one as a feature:
\scalebox{0.85}{\parbox{\columnwidth}{%
\begin{gather*}
Ochiai = \frac{e_f}{\sqrt{N_f \cdot (e_f + e_p)}} \\
DStar^{2} = \frac{e_f^{2}}{e_p + (N_f - e_f)} \\
Tarantula = 1 - \frac{h}{h + e_f / N_f} \  where \  h = \frac{e_p}{N_p}
\end{gather*}
}}

For all divisions, we add a small $\epsilon$ to the denominator to avoid division by zero.

\item[Incoming/outgoing paths (control flow)]

We use the line number to calculate some \emph{very simple} control flow features. Remember that our traces contain pairs $(l, l')$ of the current and previously executed line number. We calculate for each line $l$ the minimum, maximum, mean and median difference $l - l'$. If control flow is linear (i.e., line $l$ is always executed right after $l'$) these features will all evaluate to 1; however, if non-linear control flow occurs at these lines, the difference between $l$ and $l'$ will be larger than 1 (if we have a forward jump from $l'$ to $l$) or even negative (for backward jumps). Next, we calculate the number of outgoing and incoming \enquote{paths} by counting for a line $l$ the number of different $l'$ and vice versa. These two features represent the control flow \enquote{fan-in} and \enquote{fan-out} for a line; put simply, they approximate how non-linear control flow is at a particular line. Of course, we calculate two different versions for each of these features, one for passing and one for failing tests.

\item[Control flow keywords (lexical)] 
Finally, we also include simple lexical features. For each line $l$ we match a simple regular expression against the corresponding code line to determine if it contains control flow-related Python keywords such as \texttt{if}, \texttt{else}, \texttt{for}, \texttt{while} and so on. This results in a single boolean feature for each keyword.

\end{description}

\subsubsection{Calculating feature windows}

So far, we have calculated features on a per-line basis. That is, for each line with line number $l$ in a program we have a feature vector  $\mathbf{x_l} =  (x_{1}^{l},x_{2}^{l},\dotsc, x_{n}^{l})$. We could fit a statistical model to predict a suspiciousness score for each such vector. However, we conjecture that neighboring lines affect the suspiciousness score of a line. 
Consequently, we combine \emph{neighboring} feature vectors into a single feature vector. This is done by sliding a window of size $w$ over each line vector (in ascending order of line numbers). This way, for a line $l$, which we call the \emph{focal line}, we obtain a \emph{contextualized} feature vector
\scalebox{0.85}{\parbox{\columnwidth}{%
\begin{align*}
 \mathbf{x_{l}^{+}} =\ & (x_{1}^{l-\floor{w/2}},\dotsc,x_{n}^{l-\floor{w/2}}, \\
					  & \cdots, \\
					  & x_{1}^{l},\dotsc,x_{n}^{l},\\
				      & \cdots, \\
                      & x_{1}^{l+\floor{w/2}},\dotsc,x_{n}^{l+\floor{w/2}})
\end{align*}}}\\
where $x^{l}_{i}$ denotes the $i$\textsuperscript{th} feature of line $l$.
We introduce padding lines at the beginning and at the end such that neighboring line vectors are also defined for the first and last lines.
The idea of combining neighboring features is somewhat reminiscent of the concept of local context in neural program repair~\citep{prennerOutContextHow2024}.
For each spectral feature, we also include the maximum and minimum value within the window as additional feature.

\subsubsection{Training}

We frame the localization problem as a binary classification task. Buggy lines $l$ will be assigned the positive class ($y_l = 1$), all other lines the negative class ($y_l = 0$). We assign classes based the ground truth patch of insertions and deletions (part of RunBugRun). In particular, for insertions of new lines we use a heuristic that defines the closest previous line in the buggy program as buggy. Of our roughly 24,000 bugs we use 90\,\% for training and the remaining 10\,\% for validation (i.e., as validation set). Note that we call this validation set instead of test set as we do \enquote{introspective} analysis (e.g., feature importance) on this data, which, strictly speaking, one is not supposed to do on actual test sets.  Then, we train a gradient boosting machine on pairs of contextualized vectors and class labels  $(\mathbf{x_l^+}, y_l)$. For this, we use the LightGBM\footnote{\url{https://lightgbm.readthedocs.io/en/stable/}} framework which, in early experiments, performed slightly better than CatBoost\footnote{\url{https://catboost.ai}}. Likewise, training under the LambdaRank~\citep{burgesRankNetLambdaRankLambdaMART} objective performed worse than our windowed vectors and was thus not pursued further.

The class distribution is heavily skewed towards the negative class (roughly by a factor of 10).
We try resampling using SOMTE-ENN~\citep{batistaStudyBehaviorSeveral2004} as implemented by the Imbalanced-learn framework~\citep{JMLR:v18:16-365} without much success. Resampling has, however, been successfully applied in FL in previous work~\citep{zhangImprovingDeeplearningbasedFault2021}.

\subsubsection{Evaluation}
We use the probability assigned to the positive (i.e., buggy) class as a score for suspiciousness. For each bug we rank all lines in descending order of suspiciousness. Ties are broken by assigning the lower rank to the first line in line number order, this way no rank is shared by multiple lines.
In line with previous work~\citep{liFaultLocalizationCode2021, liDeepFLIntegratingMultiple2019, qianAGFLGraphConvolutional2021, qianGNet4FLEffectiveFault2023, zhangContextAwareNeuralFault2023} we use the following three evaluation measures where $\hat{r}_l$ is the predicted rank for line $l$ and $\hat{r}^{*}$ is the lowest rank $\hat{r}_l$ with $y_l = 1$, that is the lowest rank assigned to any buggy line for a specific bug. Note that the lowest possible rank is one.

\begin{description}
\item[Mean First Rank (MFR)] We calculate MFR as the mean of all $\hat{r}^{*}$ over all bugs under evaluation.
\item[Mean Average Rank (MAR)]
The calculation of the MAR measure is similar: for each bug under evaluation we calculate the average of \emph{all} ranks $\hat{r}_l$ with $y_l = 1$. Then we calculate the mean of such averages over all bugs under evaluation.
\item[Top-$\mathbf{N}$]
Top-$N$ is calculated as the portion of bugs under evaluation where $\hat{r}^{*} \leq N$. We calculate this measure for $N=1$, $3$ and $5$.
\end{description}

\paragraph*{Baselines}
We use previously defined Ochiai, DStar\textsuperscript{2} and Tarantula ranking metrics as baselines. An additional random baseline assigns suspiciousness scores randomly.

\paragraph*{QuixBugs}
QuixBugs~\citep{linQuixBugsMultilingualProgram2017} is a collection of 40 bugs in simple algorithm implementations written in Java and Python. In addition to our validation set we also evaluate a subset of 27 Python bugs taken from QuixBugs. In particular, we exclude bugs that have no passing tests. Some tests cause timeouts or errors. Since our training data contains only passing or failing test runs (i.e., the program either outputs a correct or incorrect result) we simply treat errors and timeouts as test failures. Including bugs with errors and timeouts in the training set might further improve performance on QuixBugs. Test executions are traced and processed in the same way as described earlier.

\section{Results}\label{sec:results}

We find that our model considerably outperforms traditional SBFL metrics on our validation set and, by a smaller margin, also on QuixBugs (see Table~\ref{tab:results}). Note that for QuixBugs, MAR and MFR are identical as the selected bugs only have a single buggy line. Results in Table~\ref{tab:results} were obtained using all features described in Section~\ref{sec:methodology} and a window size $w$ of 3 (i.e., the preceding and the succeeding lines are included). We analyze the contribution of different feature types as well as the impact of the window size in Section~\ref{sec:ablation}. 

Overall, MAR and MFR are lower (better) on QuixBugs. An explanation for this might be that the number of traced (i.e., executed) lines is much lower for the bugs in QuixBugs (average 7.6) than it is for bugs in RunBugRun (average 13.5). Also, bugs in QuixBugs only have a single buggy line. In contrast, for Top-$N$, our model performs significantly worse on QuixBugs; possibly because the validation set's distribution is much closer to the training set than QuixBugs.

\begin{table}[htbp]
\caption{Evaluation results with a window size of $w = 3$ }
\begin{center}
\begin{tabular}{ccrcccc}
& & \textbf{MAR} & \textbf{MFR} & \textbf{Top-1} & \textbf{Top-3} & \textbf{Top-5} \\
\toprule
\multirow{4}{*}{\makecell{Validation\\Set}}
& DStar\textsuperscript{2} & 6.57 & 5.43 & 23.4\% & 49.4\% & 66.7\% \\
& Ochiai & 6.58 & 5.44 & 23.4\% & 49.3\% & 66.7\% \\
& \textbf{Ours} & \textbf{3.69} & \textbf{2.66} & \textbf{56.4\%} & \textbf{79.9\%} & \textbf{88.2\% }\\
& Random & 7.27 & 5.97 & 16.9\% & 43.8\% & 62.3\% \\
& Tarantula & 6.61 & 5.45 & 23.6\% & 49.2\% & 66.1\% \\
\midrule
\multirow{4}{*}{QuixBugs}
& DStar\textsuperscript{2} & 3.70 & 3.70 & \textbf{29.6\,\%} & 55.6\,\% & \textbf{77.8\,\%}\\
& Ochiai & 3.74 & 3.74 & \textbf{29.6\,\%} & 51.9\,\% & \textbf{77.8\,\%}\\
& \textbf{Ours} & \textbf{3.33} & \textbf{3.33} & \textbf{29.6\,\%} & \textbf{63.0\,\%} & 74.1\,\%\\
& Random & 4.04 & 4.04 & 7.4\,\% & 51.9\,\% & \textbf{77.8\,\%}\\
& Tarantula & 3.78 & 3.78 & \textbf{29.6\,\%} & 51.9\,\% & \textbf{77.8\,\%}\\
\bottomrule
\end{tabular}
\label{tab:results}
\end{center}
\end{table}

\subsection{Ablation Study}\label{sec:ablation}

An interesting question is to what extent different features contribute to the performance of our model. Here, we provide a brief ablation study. 
We first show performance impact of different groups of features (e.g., lexical, spectral etc.). Next, we also look at how training set and window size affect performance. Finally, we look at the most important features as reported by the LightGBM framework.

\paragraph*{Feature Groups}
{
\setlength{\tabcolsep}{5pt}
\begin{table}[htbp]
\caption{Results of ablation experiment.}
\begin{center}
\begin{tabular}{rccccc}
& \textbf{MAR} & \textbf{MFR} & \textbf{Top-1} & \textbf{Top-3} & \textbf{Top-5} \\
\toprule
\emph{No} lexical feat. & \textbf{3.66} & \textbf{2.66} & \textbf{56.6\,\%} & \textbf{80.2\,\%} & 88.5\,\%\\
All feat. & 3.69 & 2.66 & 56.4\,\% & 79.9\,\% & 88.2\,\%\\
\emph{No} formulae feat. & 3.70 & 2.70 & 55.4\,\% & 80.1\,\% & \textbf{89.0\,\%}\\
\emph{No} spectral feat. & 3.84 & 2.83 & 54.2\,\% & 78.1\,\% & 87.2\,\%\\
\emph{No} path feat. & 3.95 & 2.90 & 54.4\,\% & 78.5\,\% & 87.0\,\%\\
\emph{No} spec.+formulae feat. & 4.48 & 3.41 & 44.5\,\% & 71.9\,\% & 83.0\,\%\\
\bottomrule
\end{tabular}
\label{tab:abl}
\end{center}
\end{table}
}

We group features into four groups, namely,
\begin{inparaenum}[1)]
\item spectral features, 
\item SBFL formulae features,
\item control flow (in/out paths) features and
\item lexical features.
\end{inparaenum} Note that SBFL formulae, while also encoding spectral information, are placed in a separate group.  
Then, for each such group, we train and evaluate a classifier omitting features belonging to that group. Training and evaluation is repeated three times per group to obtain more stable results.
As can be seen in Table~\ref{tab:abl}, including lexical features slightly lowers performance. Not surprisingly, omitting all spectral features, that is formulae score features as well as \enquote{direct} spectral features, greatly reduces performance. Still, path features are more important than either formulae or direct spectral features alone.

\paragraph*{Window size}
\begin{figure}[htbp]
\centerline{\includegraphics{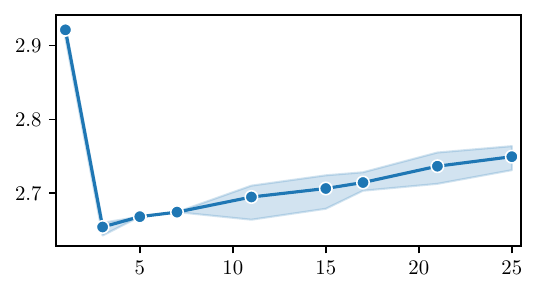}}
\caption{MFR as a function of window size (x). The lowest MFR is obtained with a window size of three. Error bands show a 95\% CI over three training runs.}
\label{fig:window-size}
\end{figure}

Figure~\ref{fig:window-size} shows how window size $w$ affects performance on our validation set. We see a pronounced drop (i.e. improvement) in MFR for $w > 1$ and a steady performance decrease for $w > 3$. In other words, we obtain the best performance for $w = 3$, that is, a single context line preceding and following the focal line.

\paragraph*{Training set size}
\begin{figure}[htbp]
\centerline{\includegraphics{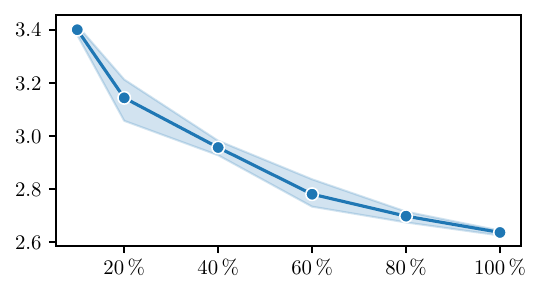}}
\caption{MFR as a function of the training set size where 100\,\% uses the entire training data of roughly 22'000 samples (buggy programs). A window size is of $w = 3$ is used here. Error bands show a 95\% CI over three training runs.}
\label{fig:training-size}
\end{figure}

We evaluate the effect of training set size by training with only a fraction of the available training data. As shown in Figure~\ref{fig:training-size}, MFR is decreasing (improving) steadily as the number of training samples is increased. While the curve starts to flatten out, we do not observe a saturation effect and more data could possibly further improve performance.

\paragraph*{Feature Importance}

{
\begin{table}[htbp]
\caption{Most important features by gain ($w = 3$)}
\begin{center}
\begin{tabular}{rcc}
\textbf{Feature} & \textbf{Gain Score} & \textbf{Rel.}  \\
\toprule
\texttt{exec\_pass\_norm0} & 49,147 & 1.00\\
\texttt{tarantula0} & 25,623 & 0.52\\
\texttt{exec\_pass\_norm\_max} & 16,470 & 0.34\\
\texttt{exec\_failed\_norm\_min} & 16,412 & 0.33\\
\texttt{num\_paths\_out\_fail1} & 16,190 & 0.33\\
\bottomrule
\end{tabular}
\label{tab:feat}
\end{center}
\end{table}
}

{
\begin{table}[htbp]
\caption{Feature importance by window level ($w = 3$)}
\begin{center}
\begin{tabular}{rcc}
\textbf{Level} & \textbf{Avg. Gain Score} & \textbf{Rel.}  \\
\toprule
Max & 7,248 & 1.00\\
Focal Line & 6,190 & 0.85\\
Min & 5,727 & 0.79\\
Succeeding Line & 4,939 & 0.68\\
Preceding Line & 3,868 & 0.53\\
\bottomrule
\end{tabular}
\label{tab:feat-win}
\end{center}
\end{table}
}

Being based on decision trees, GBMs are to some degree \emph{explainable}. In particular, LightGBM can report feature importance scores after a training run. Table~\ref{tab:feat} list the most important features by gain score. This score represents a value proportional to the decrease in training loss when using the corresponding feature for splitting. Note that due to multicollinearity in our data, care must be taken when interpreting these scores. 

Still looking at table ~\ref{tab:feat}, we see that the \emph{normalized} number of line executions under passing tests (\texttt{exec\_pass\_norm0}) is reported as the most important feature. The zero at the end of the feature name indicates that this feature belongs to the focal line and not to a neighboring context line (i.e., features for the succeeding line would end in \texttt{1} and for the preceding line in \texttt{-1}). It is perhaps worth noting that the first \emph{non}-normalized execution count feature only appears in 12\textsuperscript{th} place (we only show top 5 in Table~\ref{tab:feat}) after 8 normalized versions of such features before it. 

The Tarantula ranking score feature, again of the focal line, is listed next, followed by two more execution count features. The latter represent the minimum and maximum feature value inside the selected window and end with \texttt{min} and \texttt{max}, respectively. Last listed is \texttt{num\_paths\_out\_fail1}, the number of outgoing paths of the line following the focal line under failing tests.

Of note, we can see that there is a wide margin between the first and second ranked features with the former being almost twice as important as the latter. Also, the ranking of the Tarantula feature should be taken with a grain of salt as there is a strong multicollinearity with the other two ranking score features (i.e., Ochiai and DStar).

\paragraph*{Window level importance}
As we have seen, features can have different positions or levels within the window. With a window size of $w = 3$, we have, beside the focal line, a preceding line and a succeeding line, as well as features for the minimum and maximum in the window. Grouping and averaging feature scores accordingly, we can estimate the importance of these levels. As can be seen from Table~\ref{tab:feat-win}, taken together, maximum in-window features have the highest gain scores, followed by features of the focal line; features relating to the succeeding line are more important than those of the preceding line.

\paragraph*{Summary of results}
In short, our simple model performs significantly better than simple SBFL formulae. Including features of the preceding and succeeding lines improves performance, however, further increasing the window size leads to a  gradual performance drop. Performance also depends on the number of training samples; even our relatively large training set does not seem to fully saturate our model. Ablation shows that spectral features are most important, but also path features seem to have significant impact. On the other hand, our simple lexical features seem to have a negative effect or, at best, no effect at all.

\section{Discussion}\label{sec:discussion}

\paragraph*{Line numbers can go a long way}
 As mentioned earlier, all of our features can be derived from execution traces. In particular, a sequence of executed line numbers is sufficient for the approach presented in this paper. Once obtained such as sequence, coverage and count spectra are trivial to calculate. We can even capture rudimentary control flow information simply by looking, for a specific line $l$, at its preceding (or alternatively, succeeding) line numbers. Computed features can be fit using efficient GBM models within minutes on modern consumer-grade hardware requiring no GPU.

\paragraph*{Complementing features}
Our ablation study shows that features of different types or from different spectra complement each other well. For instance, the top five most important features contain coverage, count spectra as well as control flow spectra. While this is not a new insight and many hybrid approaches have been proposed in the literature (e.g., DeepFL~\citep{liDeepFLIntegratingMultiple2019}), it seems worth noting that complementary features can be obtained from the same information source, namely simple execution traces with line numbers.

\paragraph*{Normalization}
Our results suggest that normalization is crucial for count spectra features.
This is not surprising, as absolute counts vary widely across programs; normalization makes execution features somewhat comparable. We use a very simple normalization scheme, dividing by the total number of executions. Future work may investigate better ways of doing this normalization. Finally, normalization may also provide a way to incorporate time (or step count) features similar to \enquote{time percentages} devised by \citet{yilmazTimeWillTell2008}. 

\paragraph*{Large windows don't help}
While using windows greatly improves localization accuracy, somewhat surprisingly, windows larger than $w = 3$ have a detrimental effect on performance. More work is needed to  understand why this is and at this point we can only offer speculation. For one, the optimal window size might be proportional to the lengths of programs. As the programs used in this work are very short, larger window might not take full effect. Also, windowing incurs padding and larger windows require more padding. Features in padded areas must be filled with dummy values (we use $-1$) which might pose some difficulty to our model.

\paragraph*{Data is important}
To our surprise, the curve in Figure~\ref{fig:training-size} does not flatten out completely. This indicates that the available data is not fully saturating our model and that using more training data might yield further performance increases. This also emphasizes the importance of data in current and future fault localization systems. However, while we used small programs from programming contests, which are relatively easy to come by, it is not clear how thousands of execution traces can be obtained for real-world programs.

\section{Limitations \& Threads to Validity}

\paragraph*{Bugs}
Despite our utmost attention, we cannot fully exclude the possibility of bugs or other mistakes in our
scripts or in our analysis.

\paragraph*{Python only}
So far, this work is limited to Python programs. Python was mainly chosen because of its \texttt{sys.settrace} feature which greatly facilitates tracing. Similar features can be found in other interpreted languages (e.g. Ruby\footnote{\url{https://github.com/ruby/tracer?tab=readme-ov-file#linetracer}}). However, for compiled languages,
 obtaining execution traces may be considerably more challenging and may require instrumenting byte code (e.g., JVM) or transforming intermediate representations (e.g., LLVM/Clang). 

\paragraph*{Real-world software}
Our experiments use small programs from programming contests. Such programs are not representative of larger, real-world software projects, as for instance found in Defects4J~\citep{justDefects4JDatabaseExisting2014}.
Given that our results show a performance drop on QuixBugs (whose bugs are somewhat similar in style to RunBugRun), it is very unlikely that a model trained on such small programs would perform well on code of larger projects.
For further experiments in this direction we would need execution traces of large, real-world projects which in turn would require a large dataset (e.g., $> 10,000$ samples) of executable bugs in real-world projects.
Unfortunately, as of writing, we are not aware of any such dataset. 

\paragraph*{Baseline}
We use simple SBFL formulae as baselines. While such formula are commonly used as baselines, much stronger FL systems have been devised in the literature. However, the two most commonly employed production-ready fault localization tools, GZoltar~\citep{camposGZoltarEclipsePlugin2012} and FLACOCO~\citep{silvaFLACOCOFaultLocalization2023}, both rely on the coverage spectrum and the Ochiai ranking formula and should thus not be very far from our baselines. Moreover, our work emphasizes on simplicity and a comparison with, say, a large language model or a system employing advanced program analysis would not be an apples-to-apples comparison. Finally, we could not find any detailed FL evaluation results on the Python version of QuixBugs in the literature.

\section{Conclusions}\label{sec:conclusions}

In this work we presented a KISS (keep it simple and straightforward) approach to fault localization that leverages execution traces. We described simple ways to obtain features belonging to multiple spectra from such traces and how a machine learning model can be trained and used as a bug localizer. Our evaluation results show that our approach, although simple, clearly outperforms SBFL coverage formulae. 

We hope that some of the presented ideas may inspire authors of existing FL tools. As we have shown, the step from pure coverage SBFL to a multi-spectral approach does not have to be very big. All that is necessary is the sequence of executed lines (e.g. line numbers). Even the transition from boolean coverage matrices to normalized count matrices may already lead to an appreciable improvement.

\paragraph*{Acknowledgments}
This study has received financial support from the French State in the framework of the Investments for the Future programme IdEx université de Bordeaux.

\printbibliography{}

\end{document}